\documentclass{article} 
\usepackage[margin=1in]{geometry}
\usepackage{amsmath,amsfonts,amssymb}
\usepackage{graphicx,wrapfig}
\usepackage[hyphens]{xurl}
\usepackage{hyperref}
\usepackage{authblk}
\newcommand{\old}[1]{{}}
\providecommand{\keywords}[1]
{
  \small	
  \textbf{\textit{Keywords---}} #1
}
\title{
GAN with Skip Patch Discriminator for Biological Electron Microscopy Image Generation}

\author{Nishith Ranjon Roy$^{1}$,
Nailah Rawnaq$^{1}$,
Tulin Kaman$^{1}$ \\ 
\small $^{1}$ Department of Mathematical Sciences, 
University of Arkansas, Fayetteville, AR 72701, USA
}
\date{} 

\begin{document}
\maketitle

\begin{abstract}
Generating realistic electron microscopy (EM) images has been a challenging problem due to their complex global and local structures. Isola et al. proposed pix2pix, a conditional Generative Adversarial Network (GAN), for the general purpose of image-to-image translation; which fails to generate realistic EM images. We propose a new architecture for the discriminator in the GAN providing access to multiple patch sizes using skip patches and generating realistic EM images.
\end{abstract}



\section{Introduction}
In classical GAN, the adversarial modeling framework consists of a generator $G$ and a discrimination $D$ model. A generator model mimics the data distribution and a discrimination model determines if an observation is from generated or real data~\cite{goodfellow2014generative}. The pix2pix model~\cite{isola2018imagetoimage} is suitable for image-to-image translation of real-life objects like cars, roads, and buildings but not for EM images.
The reason is that EM images are high-resolution microscopy images containing complex structures consisting of intricate, asymmetric and sparse patterns which are different from any regular objects. In addition, the limited data availability for EM images is a concern to train a neural network. 
We address the challenges and propose a new approach for training GAN models in settings with limited data availability and the presence of both global and local structures for generating realistic biological images.

\subsection{Dataset (Drosophila)}
\begin{figure}[!t]
    \centering
    \includegraphics[scale=.3]{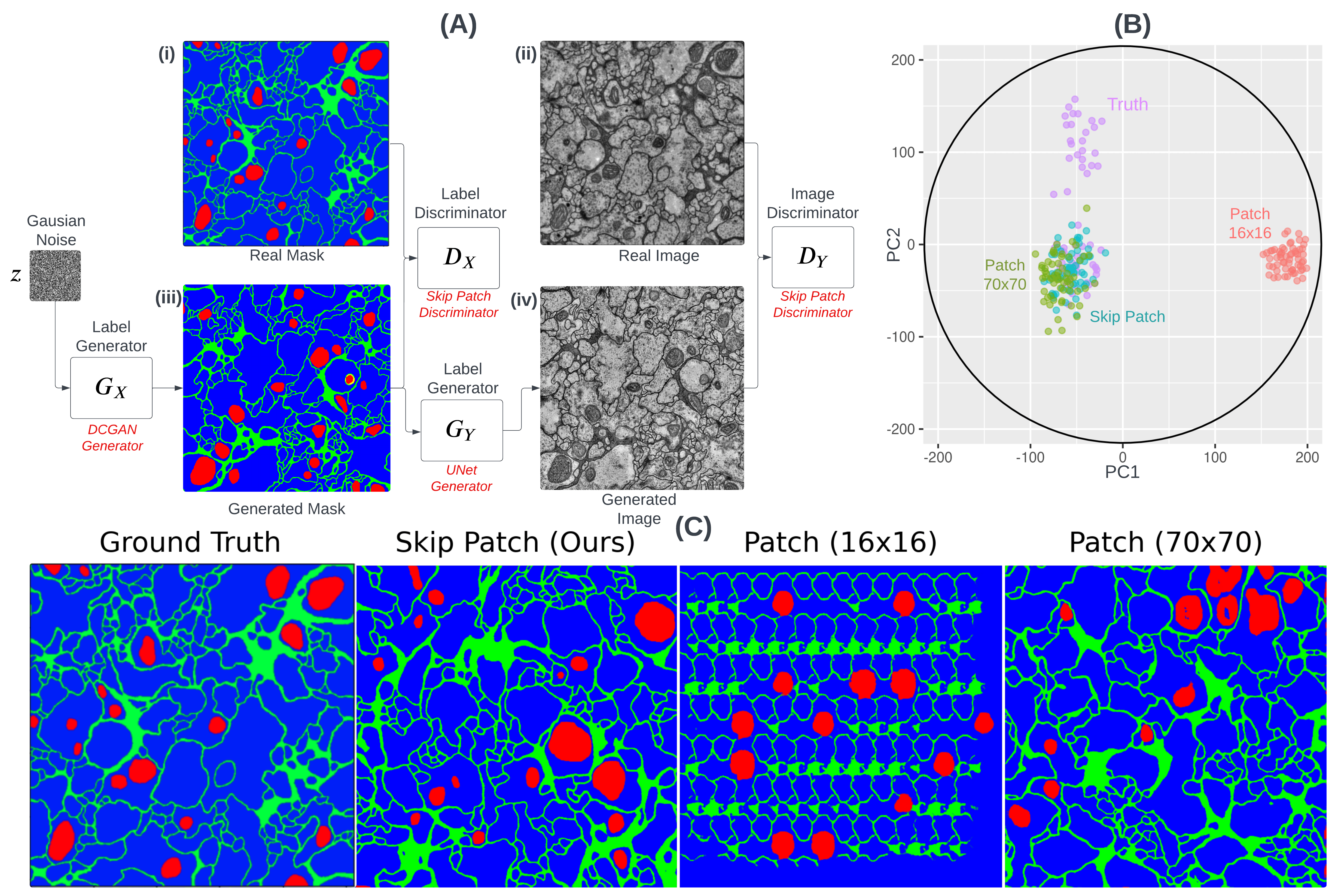}
    \caption{(A) The process of generating masks ($X$) and EM images ($Y$). The generator and discriminator are denoted by $G$ and $D$ respectively. The first stage of this process involves generating masks (iii) from random noise ($z$), while the second stage involves using these generated masks to produce EM images (iv). (B) The principal component analysis (PCA) plot shows the similarity among the distribution of the generated mask with the real mask. The red points comparatively far on the right side indicate that the generated mask with $16\times 16$ patch does not show any alignment with the true mask. On the contrary, the mixture in the center of three different points suggests that the generated mask with $70\times70$ patch, skip patch (ours), and real mask have approximately the same type of distribution. (C) The four images show the real and generated masks via different methods. A fixed 6000 training iterations (or epochs) has been performed for all three methods.}
    \label{fig:figure1_comb}
\end{figure}

We run our studies on the Drosophila dataset~\cite{gerhard2013dataset} which consists of 
$20$ ($1024\times 1024$ resolution) images of the mitochondria and cell membrane masks (see Fig.~\ref{fig:figure1_comb}-A.i), and EM images (see Fig.~\ref{fig:figure1_comb}-A.ii).

\section{Methodology}
The process of generating EM images consists of two steps. The first step is to create the mitochondria and cell membrane masks, and the second step is to create EM images of tissue organization. 
In this paper, we present a new architecture for the discriminator in training GAN models. 
In both steps of the GAN training, the skip patch discriminator (proposed here) and the patch discriminator (initially proposed by \cite{isola2018imagetoimage}) are used and compared. 
In Fig.~\ref{fig:figure1_comb}-A, we show the overall process of generating masks and images.

\subsection{Skip patch discriminator}
\begin{figure}[!t]
    \centering
    \includegraphics[scale=.6]{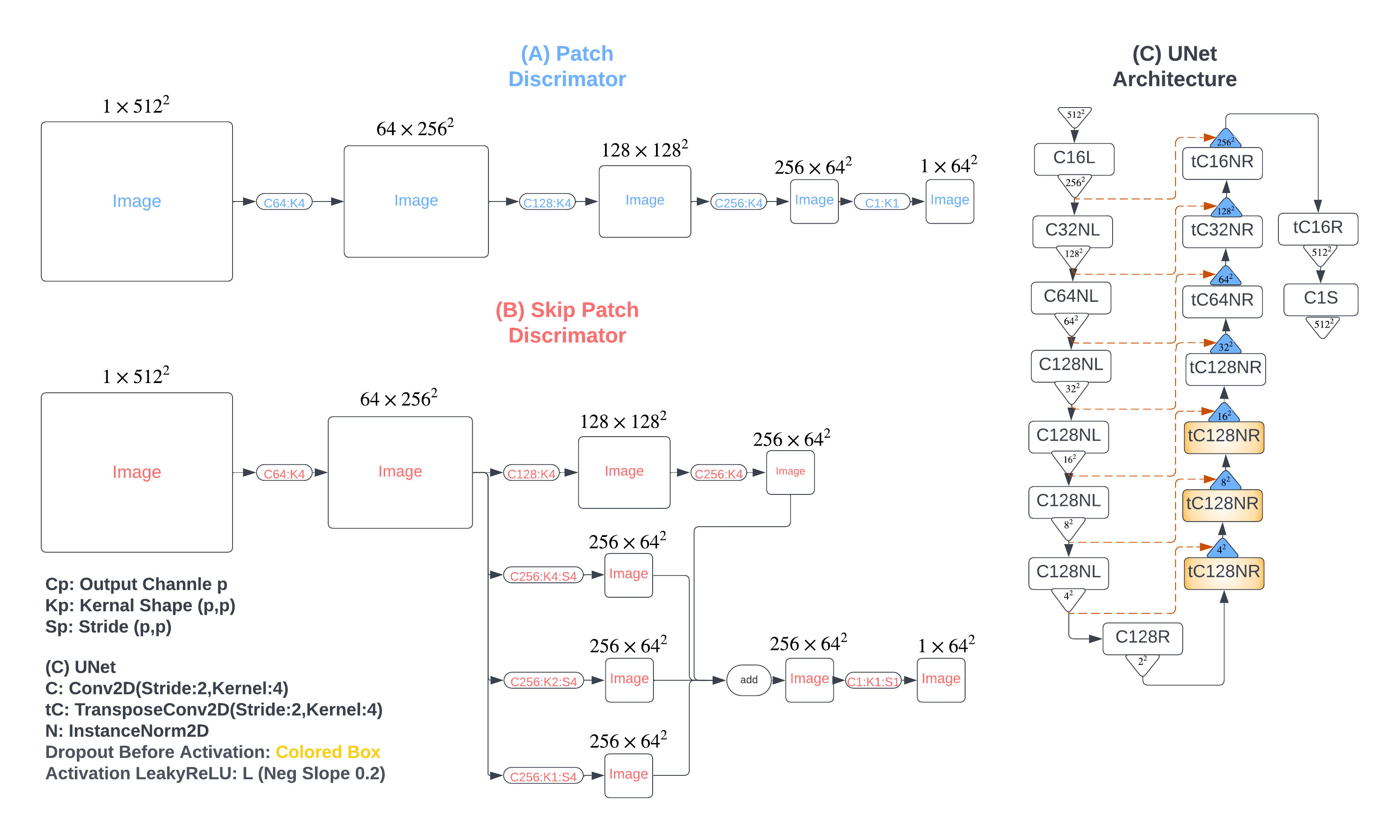}
    \caption{(A) The discriminator architecture proposed by \cite{goodfellow2014generative}. 
    (B) The new discriminator architecture proposed in this paper. 
    (C) U-Net architecture for the generator in the conditional GAN (cGAN) EM images generation part.}
    \label{fig:model_arch}
\end{figure}

The patch-based discriminator lacks the capability of simultaneously accessing both the global and local structures of the generated image. The discriminator outputs a matrix where each element of that matrix is calculated by applying convolution on the input image/tensor. The backtracking of the calculation for one pixel in the final layer of the patch-based discriminator is responsible for $70\times 70$ pixels. Fig.~\ref{fig:model_arch}-A shows the $70\times 70$ patches in the input tensor/image which allows the discriminator to evaluate the generated image on a $70\times 70$ window. As shown in Fig.~\ref{fig:figure1_comb}-C, the patch size $70\times 70$ captures the global structure but fails to capture the local structure and results in holes inside some of the mitochondria. Cutting down the depth of the number of layers from four to two reduces the patch size from $70\times 70$ to $16\times 16$ which captures the local structure but fails to capture the global structure and results in unrealistic symmetrical repetitive shapes.

Therefore, we propose a skip patch discriminator (see Fig.~\ref{fig:model_arch}-B), where, due to the skip connections, each pixel of the final layer's output tensor/matrix has access to the patch sizes of $16\times 16$, $20 \times 20$, $32\times 32$, and $70\times 70$. Consequently, this discriminator is now capable of detecting trends, symmetry, and asymmetry across a range of different window sizes, from a minimum of $16\times 16$ to a maximum of $70\times 70$ parches. See Fig.~\ref{fig:figure1_comb}-C ``skip patch'' and Fig.~\ref{fig:result}-B ``cGAN~S.Patch''.

\subsection{Mitochondria and cell membrane mask (Fig.~\ref{fig:figure1_comb}-A.i)}
\textbf{Generator}: The generator is defined as 
$G: N(0,1)^{8\times8\times8} \to (0,1)^{3^*\times512\times512}$ where $3^*$ represents the number of channels corresponding to the mitochondria (red) and cell membrane (green) mask, and the background (blue), and $N(0,1)$ is a Gaussian distribution with a mean $0$ and 
a standard deviation $1$. The architecture of the model is as follows: $C256NA \to C256NA \to C128NA \to C64NA \to C32NA \to C3A$. Here, C${i}$ denotes the transposed convolution with the number of output channels $i$ (kernel 4, stride 2), followed by the instance normalization (N) and ReLU activation (A). In the final layer, softmax activation is used without a normalization layer.

\textbf{Discriminator}: A skip patch discriminator resulted in a realistic-looking mask compared to a patch discriminator (see Fig.~\ref{fig:figure1_comb}-C). All of these models have been trained on a fixed number (6000) of epochs and the setup with skip patch discriminator starts to converge within 50\% of the total epoch. Fig.~\ref{fig:model_arch}-B shows the new architecture of the network where the discriminator is defined as $D:(0,1)^{3\times512\times512}\to\{0,1\}^{64\times64}$.

\textbf{Training}: The tensor containing the information of the mask was resized down to $512 \times 512$ resolution. For augmentation, random vertical and horizontal flipping were applied. Additionally, cropping with 90\% of the area (keeping the aspect ratio fixed) was applied to the mask tensor to address overfitting. 
The objective function of GAN for mask generation is
\begin{equation*}
\min_{(G)}~\max_{(D)}~V(G,D) = {{\cal L}(\text{Prob=}D_y(G_x(z)),\text{Target=}\{0\}^{64\times64}) + {\cal L}(\text{Prob=}D_y(X), \text{Target=}\{1\}^{64\times64})}
\label{eq:objGAN}
\end{equation*}
where $z$ is the random noise, $X$ is the mask, $Y$ is the image and ${\cal L}$ is the binary cross entropy.

\subsection{EM Image of cellular structure (Fig.~\ref{fig:figure1_comb}-A.ii)}
\textbf{Generator}: The generator is defined as $G:\{0,1\}^{3\times512\times512}\to(0,1)^{1\times512\times512}$ is a U-Net based architecture used by Isola {\it et al.}~\cite{isola2018imagetoimage}. Due to the tendency to show artifacts in the generated images \cite{zhu2020unpaired} changed the batch normalization layer with instance normalization (see Fig.~\ref{fig:model_arch}-C).

\textbf{Discriminator}: In total three different conditional GAN models have been fit. Fig.~\ref{fig:result}-PCA plot shows that all three models have the distribution of generated images as the real images. However, after observing the generated images the conclusion can be made that the cGAN model with skip patch (ours) discriminator provides a better result. A closer examination of the images reveals the artifact inside the generated mitochondria images by the patch discriminator. Besides, the model with skip patch discriminator converges 50\% faster than the patch discriminator. All models have been trained for a 1500 fixed epoch size. Since the 3 channel of the mask is being concatenated with the image, the discriminator is defined as $D:(0,1)^{4\times512\times512}\to\{0,1\}^{64\times64}$

\textbf{Training}: As the dataset is reasonably small with only 20 examples it is quite hard to retain the model from overfitting. 
As a remedy, a comparatively lower parameterized generator architecture has been used with an initial number of channels equal to 16 (see Fig.~\ref{fig:model_arch}-C). Additionally, for the augmentation random vertical flipping, horizontal flipping, and cropping with 98\% of the area (keeping the aspect ratio fixed) were used. The objective of cGAN is 
\begin{equation*}
\arg \min_{(G)} \max_{(D)} E_{x,y}[\log~D(x,y)]+E_{x,z}[\log(1-D(x,G(x,z)))] +\lambda L_{L1}
\label{eq:obj_cGan}
\end{equation*}
where $L_{L1}$ is the $L1$ loss and $\lambda$ is the weight for the $L1$ loss.

\section{Results and conclusions}
\begin{figure}[!t]
    \centering
    \includegraphics[scale=.55]{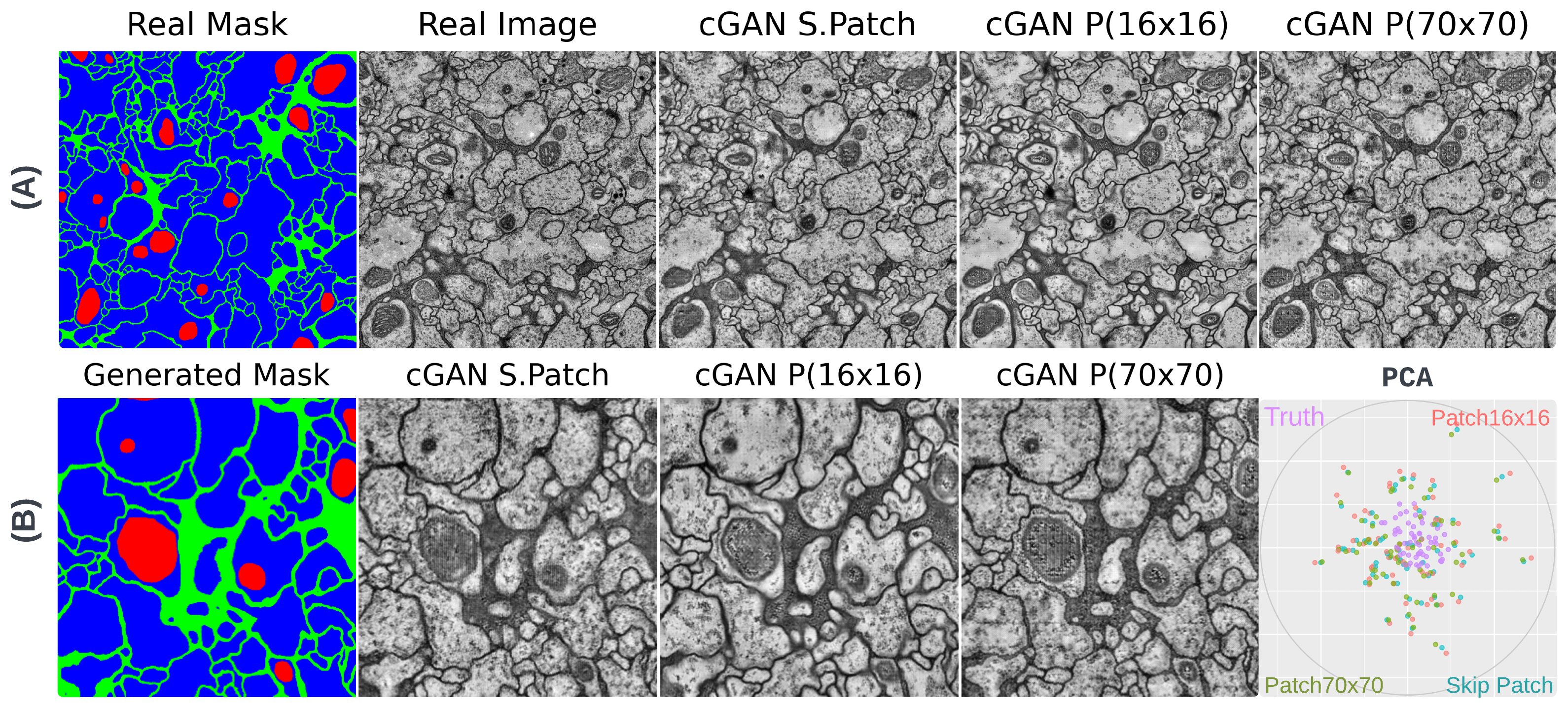}
    \caption{(A) Outcome of the conditional GAN (cGAN) models with the training data where S.Patch, P(16x16) and P(70x70) correspond to the skip-patch, $16\times 16$ patch and $70\times 70$ patch discriminators respectively. (B) The outcome of the cGAN models with the newly generated mask (zoomed). All of the points mix evenly in the PCA plot of the generated images vs the real image. Hence according to PCA, the distribution of the generated images with all of the three methods has a distribution close to the real EM images.}
    \label{fig:result}
\end{figure}

For the generation of EM images, the PCA method suggests that any of the options (patch $16\times 16$, patch $70\times 70$, or skip patch discriminator) will have a similar image distribution as the real EM images. However, a closer inspection of the generated images will reveal that the discriminator with a $16\times 16$ patch produces blurry images, and a discriminator with a $70\times 70$ patch produces images with a subtle square artifact inside all of the mitochondria. Using a skip patch discriminator provides a better result than using a patch discriminator even though the empirical result of PCA suggests that all models are equally good (see Fig~\ref{fig:result}-B). Also, the cGAN model with skip patch discriminator converges approximately with a total epoch of 50\% compared to the patch discriminator cGAN. Though for unbiased comparison all models have been trained for the same number of epochs indifferent of convergence. All of the networks produce similar types of images on the training data and the new unseen data which suggests that the models have not overfitted the training data. So, for a small dataset, a cGAN with a smaller number of parameters in the generator architecture is quite helpful.

For generating the mitochondria and cell membrane mask (the main focus of our study), a patch discriminator does not perform well on both the global and local structures. Replacing the patch discriminator with the skip patch discriminator forces the generator to learn about both the global and local structures. Additionally, approximately 50\% less epoch is required to obtain the desired performance compared to the patch discriminator. Without using the skip patch discriminator it is quite impossible to generate intricate structures such as the mitochondria and cell membrane mask~(see Fig.~\ref{fig:figure1_comb}-C). Finally, when a generated image requires some global and local patterns to exhibit, a GAN with a skip-patch discriminator is always recommended.

\end{document}